\documentclass[trackchanges,twocolumn]{aastex701}

\newcommand\pulsar{PSR J2108+5055}
\newcommand{\red}[1]{\textcolor{black}{#1}}
\usepackage{mathtools}
\usepackage{siunitx}
\usepackage{amsmath}
\usepackage{tipa}
\DeclarePairedDelimiterXPP\BigOSI[2]%
  {\mathcal{O}}{(}{)}{}%
  {\SI{#1}{#2}}
\begin{document}

\title{Discovery of an Extremely Luminous Sporadic Radio Pulsar} 
\author[0000-0003-4098-5222]{Fengqiu Adam Dong}
  \affiliation{Department of Physics and Astronomy, York University, 4700 Keele Street, Toronto, ON MJ3 1P3, Canada}
  \email{fengqiu.dong@gmail.com}
\author[0000-0002-7164-9507]{Robert Main}
  \affiliation{Department of Physics, McGill University, 3600 rue University, Montr\'eal, QC H3A 2T8, Canada}
  \affiliation{Trottier Space Institute, McGill University, 3550 rue University, Montr\'eal, QC H3A 2A7, Canada}
  \email{robert.main@mcgill.ca}
\author[0009-0006-8984-9220]{Jackson D.~Taylor}
  \affiliation{Department of Physics and Astronomy, West Virginia University, PO Box 6315, Morgantown, WV 26506, USA }
  \affiliation{Center for Gravitational Waves and Cosmology, West Virginia University, Chestnut Ridge Research Building, Morgantown, WV 26505, USA}
  \email{jdt00012@mix.wvu.edu}
\author[0000-0002-3980-815X]{Shion Andrew}
  \affiliation{MIT Kavli Institute for Astrophysics and Space Research, Massachusetts Institute of Technology, 77 Massachusetts Ave, Cambridge, MA 02139, USA}
  \affiliation{Department of Physics, Massachusetts Institute of Technology, 77 Massachusetts Ave, Cambridge, MA 02139, USA}
  \email{shiona@mit.edu}
\author[0009-0007-0757-9800]{Alyssa Cassity}
  \affiliation{Department of Physics and Astronomy, University of British Columbia, 6224 Agricultural Road, Vancouver, BC V6T 1Z1 Canada}
  \email{acassity@phas.ubc.ca}
\author[0000-0002-2878-1502]{Shami Chatterjee}
  \affiliation{Cornell Center for Astrophysics and Planetary Science, Cornell University, Ithaca, NY 14853, USA}
  \email{shami@astro.cornell.edu}
\author[0000-0002-8376-1563]{Alice P.~Curtin}
  \affiliation{Department of Physics, McGill University, 3600 rue University, Montr\'eal, QC H3A 2T8, Canada}
  \affiliation{Trottier Space Institute, McGill University, 3550 rue University, Montr\'eal, QC H3A 2A7, Canada}
  \affiliation{Anton Pannekoek Institute for Astronomy, University of Amsterdam, Science Park 904, 1098 XH Amsterdam, The Netherlands}
  \email{alice.curtin@mail.mcgill.ca}
\author[0000-0001-8384-5049]{Emmanuel Fonseca}
  \affiliation{Department of Physics and Astronomy, West Virginia University, PO Box 6315, Morgantown, WV 26506, USA }
  \affiliation{Center for Gravitational Waves and Cosmology, West Virginia University, Chestnut Ridge Research Building, Morgantown, WV 26505, USA}
  \email{emmanuel.fonseca@mail.wvu.edu}
\author[0000-0002-3382-9558]{B.~M.~Gaensler}
  \affiliation{Department of Astronomy and Astrophysics, University of California, Santa Cruz, 1156 High Street, Santa Cruz, CA 95060, USA}
  \affiliation{Dunlap Institute for Astronomy and Astrophysics, 50 St. George Street, University of Toronto, ON M5S 3H4, Canada}
  \affiliation{David A. Dunlap Department of Astronomy and Astrophysics, 50 St. George Street, University of Toronto, ON M5S 3H4, Canada}
  \email{gaensler@ucsc.edu}
\author[0000-0003-2317-1446]{Jason Hessels}
  \affiliation{Department of Physics, McGill University, 3600 rue University, Montr\'eal, QC H3A 2T8, Canada}
  \affiliation{Trottier Space Institute, McGill University, 3550 rue University, Montr\'eal, QC H3A 2A7, Canada}
  \affiliation{Anton Pannekoek Institute for Astronomy, University of Amsterdam, Science Park 904, 1098 XH Amsterdam, The Netherlands}
  \affiliation{ASTRON, Netherlands Institute for Radio Astronomy, Oude Hoogeveensedijk 4, 7991 PD Dwingeloo, The Netherlands}
  \email{jason.hessels@mcgill.ca}
\author[0000-0001-9345-0307]{Victoria Kaspi}
  \affiliation{Department of Physics, McGill University, 3600 rue University, Montr\'eal, QC H3A 2T8, Canada}
  \affiliation{Trottier Space Institute, McGill University, 3550 rue University, Montr\'eal, QC H3A 2A7, Canada}
  \email{victoria.kaspi@mcgill.ca}
\author[0009-0004-4176-0062]{Afrokk Khan}
  \affiliation{Department of Physics, McGill University, 3600 rue University, Montr\'eal, QC H3A 2T8, Canada}
  \affiliation{Trottier Space Institute, McGill University, 3550 rue University, Montr\'eal, QC H3A 2A7, Canada}
  \email{afrasiyab.khan@mcgill.ca}
\author[0000-0003-4634-5453]{Lars K\"{u}nkel}
  \affiliation{Department of Physics, McGill University, 3600 rue University, Montr\'eal, QC H3A 2T8, Canada}
  \affiliation{Trottier Space Institute, McGill University, 3550 rue University, Montr\'eal, QC H3A 2A7, Canada}
  \email{lars.kuenkel@mcgill.ca}
\author[0000-0002-5857-4264]{Mattias Lazda}
  \affiliation{Dunlap Institute for Astronomy and Astrophysics, 50 St. George Street, University of Toronto, ON M5S 3H4, Canada}
  \affiliation{David A. Dunlap Department of Astronomy and Astrophysics, 50 St. George Street, University of Toronto, ON M5S 3H4, Canada}
  \email{mattias.lazda@mail.utoronto.ca}
\author{Calvin Leung}
  \affiliation{Miller Institute for Basic Research, University of California, Berkeley, CA 94720, United States}
  \affiliation{Department of Astronomy, University of California, Berkeley, CA 94720, United States}
  \email{calvin_leung@berkeley.edu}
\author[0000-0002-4279-6946]{Kiyoshi W.~Masui}
  \affiliation{MIT Kavli Institute for Astrophysics and Space Research, Massachusetts Institute of Technology, 77 Massachusetts Ave, Cambridge, MA 02139, USA}
  \affiliation{Department of Physics, Massachusetts Institute of Technology, 77 Massachusetts Ave, Cambridge, MA 02139, USA}
  \email{kmasui@mit.edu}
\author[0000-0001-7348-6900]{Ryan Mckinven}
  \affiliation{Department of Physics, McGill University, 3600 rue University, Montr\'eal, QC H3A 2T8, Canada}
  \affiliation{Trottier Space Institute, McGill University, 3550 rue University, Montr\'eal, QC H3A 2A7, Canada}
  \email{ryan.mckinven@mcgill.ca}
\author[0000-0002-0940-6563]{Mason Ng}
  \affiliation{Department of Physics, McGill University, 3600 rue University, Montr\'eal, QC H3A 2T8, Canada}
  \affiliation{Trottier Space Institute, McGill University, 3550 rue University, Montr\'eal, QC H3A 2A7, Canada}
  \email{mason.ng@mcgill.ca}
\author[0000-0002-8897-1973]{Ayush Pandhi}
  \affiliation{Department of Physics, McGill University, 3600 rue University, Montr\'eal, QC H3A 2T8, Canada}
  \affiliation{Trottier Space Institute, McGill University, 3550 rue University, Montr\'eal, QC H3A 2A7, Canada}
  \email{ayush.pandhi@mcgill.ca}
\author[0000-0002-8912-0732]{Aaron B.~Pearlman}
  \altaffiliation{NASA Hubble Fellow.}
  \affiliation{MIT Kavli Institute for Astrophysics and Space Research, Massachusetts Institute of Technology, 77 Massachusetts Avenue, Cambridge, MA 02139, USA}
  \affiliation{Department of Physics, McGill University, 3600 rue University, Montr\'eal, QC H3A 2T8, Canada}
  \affiliation{Trottier Space Institute, McGill University, 3550 rue University, Montr\'eal, QC H3A 2A7, Canada}
  \email{aaron.b.pearlman@mit.edu}
\author[0000-0002-4795-697X]{Ziggy Pleunis}
  \affiliation{Anton Pannekoek Institute for Astronomy, University of Amsterdam, Science Park 904, 1098 XH Amsterdam, The Netherlands}
  \affiliation{ASTRON, Netherlands Institute for Radio Astronomy, Oude Hoogeveensedijk 4, 7991 PD Dwingeloo, The Netherlands}
  \email{z.pleunis@uva.nl}
\author[0000-0002-3430-7671]{Alexander W.~Pollak}
  \affiliation{SETI Institute, 339 Bernardo Ave, Suite 200 Mountain View, CA 94043, USA}
  \email{apollak@seti.org}
\author[0009-0008-2000-6959]{Sachin Pradeep E.~T.}
  \affiliation{Department of Physics, McGill University, 3600 rue University, Montr\'eal, QC H3A 2T8, Canada}
  \affiliation{Trottier Space Institute, McGill University, 3550 rue University, Montr\'eal, QC H3A 2A7, Canada}
  \email{sachin.pradeepetakkepravanthulicheri@mail.mcgill.ca}
\author[0000-0001-5799-9714]{Scott M.~Ransom}
  \affiliation{National Radio Astronomy Observatory, 520 Edgemont Rd, Charlottesville, VA 22903, USA}
  \email{sransom@nrao.edu}
\author[0000-0002-7374-7119]{Paul Scholz}
  \affiliation{Department of Physics and Astronomy, York University, 4700 Keele Street, Toronto, ON MJ3 1P3, Canada}
  \email{pscholz@yorku.ca}
\author[0000-0002-6823-2073]{Kaitlyn Shin}
  \affiliation{Cahill Center for Astronomy and Astrophysics, MC 249-17 California Institute of Technology, Pasadena CA 91125, USA}
  \email{kaitshin@caltech.edu}
\author[0000-0002-2088-3125]{Kendrick Smith}
  \affiliation{Perimeter Institute of Theoretical Physics, 31 Caroline Street North, Waterloo, ON N2L 2Y5, Canada}
  \email{kmsmith@perimeterinstitute.ca}
\author[0000-0001-9784-8670]{Ingrid Stairs}
  \affiliation{Department of Physics and Astronomy, University of British Columbia, 6224 Agricultural Road, Vancouver, BC V6T 1Z1 Canada}
  \email{stairs@astro.ubc.ca}
\newcommand{\allacks}{
We acknowledge that CHIME is located on the traditional, ancestral, and unceded territory of the Syilx/Okanagan people. We are grateful to the staff of the Dominion Radio Astrophysical Observatory, which is operated by the National Research Council of Canada. CHIME operations are funded by a grant from the NSERC Alliance Program and by support from McGill University, University of British Columbia, and University of Toronto. CHIME was funded by a grant from the Canada Foundation for Innovation (CFI) 2012 Leading Edge Fund (Project 31170) and by contributions from the provinces of British Columbia, Québec and Ontario. The CHIME/FRB Project was funded by a grant from the CFI 2015 Innovation Fund (Project 33213) and by contributions from the provinces of British Columbia and Québec, and by the Dunlap Institute for Astronomy and Astrophysics at the University of Toronto. Additional support was provided by the Canadian Institute for Advanced Research (CIFAR), the Trottier Space Institute at McGill University, and the University of British Columbia. The CHIME/FRB baseband recording system is funded in part by a CFI John R. Evans Leaders Fund award to IHS.

We acknowledge that CHIME and the \textbf{k’ni\textipa{P}atn k’l$_\smile$stk’masqt} Outrigger (KKO) are built on the traditional, ancestral, and unceded territory of the Syilx Okanagan people. \textbf{k’ni\textipa{P}atn k’l$_\smile$stk’masqt} is situated on land leased from the Imperial Metals Corporation. We are grateful to the staff of the Dominion Radio Astrophysical Observatory, which is operated by the National Research Council of Canada. CHIME operations are funded by a grant from the NSERC Alliance Program and by support from McGill University, University of British Columbia, and University of Toronto. CHIME/FRB Outriggers are funded by a grant from the Gordon \& Betty Moore Foundation. We are grateful to Robert Kirshner for early support and encouragement of the CHIME/FRB Outriggers Project, and to Dusan Pejakovic of the Moore Foundation for continued support. CHIME was funded by a grant from the Canada Foundation for Innovation (CFI) 2012 Leading Edge Fund (Project 31170) and by contributions from the provinces of British Columbia, Québec and Ontario. The CHIME/FRB Project was funded by a grant from the CFI 2015 Innovation Fund (Project 33213) and by contributions from the provinces of British Columbia and Québec, and by the Dunlap Institute for Astronomy and Astrophysics at the University of Toronto. Additional support was provided by the Canadian Institute for Advanced Research (CIFAR), the Trottier Space Institute at McGill University, and the University of British Columbia. The CHIME/FRB baseband recording system is funded in part by a CFI John R. Evans Leaders Fund award to IHS.

F.A.D is a Canadian SKA Scientist and is funded by the Government of Canada / est financé par le gouvernement du Canada. F.A.D also acknowledges prior support by the National Radio Astronomy Observatory with the Jansky Postdoctoral Fellowship.
The NASA West Virginia Space Grant Consortium, Grant~\#~80NSSC25M7079 provided individual support to J.D.T.
A.C. is supported by a UBC Four Year Fellowship
A.P.C. is a Canadian SKA Scientist and is funded by the Government of Canada / est financé par le gouvernement du Canada.
E.F. is supported by the National Science Foundation under grant AST-2407399.
The AstroFlash research group at McGill University, University of Amsterdam, ASTRON, and JIVE is supported by: a Canada Excellence Research Chair in Transient Astrophysics (CERC-2022-00009); an Advanced Grant from the European Research Council (ERC) under the European Union’s Horizon 2020 research and innovation programme (`EuroFlash’; Grant agreement No. 101098079); an NWO-Vici grant (`AstroFlash’; VI.C.192.045); an NSERC Discovery Grant (RGPIN-2025-06681); an ERC Starting Grant (`EnviroFlash’; Grant agreement No. 101223057); and an NWO-Veni grant (VI.Veni.222.295).
 V.M.K. holds the Lorne Trottier Chair in Astrophysics \& Cosmology, a Distinguished James McGill Professorship, and receives support from an NSERC Discovery grant (RGPIN 228738-13). 
M.L. acknowledges the support of the Natural Sciences and Engineering Research Council of Canada (NSERC-CGSD).
K.W.M. is supported by NSF Grant Nos. 2008031, 2510771 and holds the Adam J. Burgasser Chair in Astrophysics.
M.N. is a Fonds de Recherche du Quebec - Nature et Technologies (FRQNT) postdoctoral fellow
A.P. is a Trottier Space Institute Postdoctoral Fellow.
A.B.P.~acknowledges support by NASA through the NASA Hubble Fellowship grant \mbox{HST-HF2-51584.001-A} awarded by the Space Telescope Science Institute, which is operated by the Association of Universities for Research in Astronomy, Inc., under NASA contract \mbox{NAS5-26555}. A.B.P.~also acknowledges prior support from a Banting Fellowship, a McGill Space Institute~(MSI) Fellowship, and a Fonds de Recherche du \mbox{Qu\'ebec -- Nature} et Technologies~(FRQNT) Postdoctoral Fellowship.
The AstroFlash research group at McGill University, University of Amsterdam, ASTRON, and JIVE is supported by: a Canada Excellence Research Chair in Transient Astrophysics (CERC-2022-00009); an Advanced Grant from the European Research Council (ERC) under the European Union’s Horizon 2020 research and innovation programme (`EuroFlash’; Grant agreement No. 101098079); an NWO-Vici grant (`AstroFlash’; VI.C.192.045); an NSERC Discovery Grant (RGPIN-2025-06681); an ERC Starting Grant (`EnviroFlash’; Grant agreement No. 101223057); and an NWO-Veni grant (VI.Veni.222.295).
S.P.E.T is a Fonds de Recherche du Quebec - Nature et Technologies (FRQNT) doctoral fellow
The National Radio Astronomy Observatory is a facility of the National Science Foundation operated under cooperative agreement by Associated Universities, Inc. SMR is a CIFAR Fellow and is supported by the NSF Physics Frontiers Center award 2020265.
P.S. acknowledges the support of an NSERC Discovery Grant (RGPIN-2024-06266).
Pulsar and FRB research at UBC is funded by an NSERC Discovery Grant and by the Canadian Institute for Advanced Research.
Editting and data analysis of this manuscript was aided by LLM models
}


\begin{abstract}
We have discovered \pulsar{}, an extremely luminous but sporadic mode-changing rotating radio transient (RRAT), with a period of 0.495\,s using the Canadian Hydrogen Intensity Mapping Experiment/Fast Radio Burst instrument (CHIME/FRB). We show that the brightest pulse from \pulsar{} has a flux density of $3.1\pm1.0$\,kJy, corresponding to a spectral luminosity of $(7.5\pm2.3)\times 10^{23}$\,erg\,s$^{-1}$\,Hz$^{-1}$, assuming a dispersion-measure-derived distance of $1.6\pm0.3$\,kpc. We place an upper limit on the duty cycle of \pulsar{} at 0.002\% and measure a surface magnetic field of $5.8\times10^{11}$\,G, a characteristic age of $1.2\times10^{7}$\,yr, and a spindown luminosity of $2.2\times10^{32}$\,erg\,s$^{-1}$. Using a simple model based on the Green Bank North Celestial Cap (GBNCC) survey, we place an upper limit on the number of \pulsar{}-like sources with similar peak flux densities in the GBNCC survey area of $\sim$80 RRATs. We show that if \pulsar{} were placed at the edge of the Local Group of galaxies, the Five-Hundred-Meter Aperture Spherical Telescope (FAST) would detect its brightest pulse. The second brightest pulse has a peak flux density of $1.5\pm0.7$\,kJy, and FAST would detect such a pulse out to $\sim$1\,Mpc, covering the inner Local Group, including Andromeda. 

\end{abstract}
\section{Introduction}
A radio pulsar is a neutron star remnant which can be created via many formation channels. These include supernovae of massive stars, accretion-induced collapse, and binary mergers. The conservation of angular momentum and off-center kicks from the supernova lead to rotation periods ranging from tens of milliseconds to seconds. For known pulsars, radio emission is thought to emerge from the magnetic poles of the resulting neutron star, which are usually misaligned with the rotational axis. The misalignment causes a pulsing effect when observed from Earth. Pulse arrival times can be recorded with high precision, enabling their use as clocks \citep[e.g.][and references therein]{Lorimer:Kramer:2012}. This is especially true for millisecond pulsars \citep{Rawley:Taylor:Davis:1987}.

A subset of pulsars is known to turn off their emission, and this is an intrinsic property of the pulsar rather than an effect of the interstellar medium \citep{McLaughlin:Lyne:Lorimer:2006,Kaplan:2018,Dong:Herrera-Martin:Stairs:2024}. The most intermittent pulsars, known as rotating radio transients (RRATs)\footnote{\url{https://github.com/rratalog/rratalog}}\citep{Agarwal:Lewis:Lorimer:2026}, can only be detected by their individual pulses \citep{McLaughlin:Lyne:Lorimer:2006,Burke-Spolaor:2013}. This is in contrast to persistent pulsars, which are most often detected via folding observations at the pulsar's rotational period, and discovered using Fourier techniques \citep[e.g.,][]{Manchester:Lyne:Camillo:2001}.

Fast radio bursts (FRBs) are bright, energetic radio bursts of extragalactic origin. Their nature remains mysterious; however, multiple lines of evidence suggest they likely originate from neutron stars. These include the only Galactic FRB-like burst from a Galactic magnetar \citep{Bochenek:Ravi:Belov:2020,10.1038/s41586-020-2863-y}, pulsar-like polarization angle swings \citep{Mckinven:Bhardwaj:Eftekhari:2025}, and emission regions constrained to sizes of a few hundred kilometers \citep{Nimmo:Pleunis:Beniamini:2025}. Among neutron stars, the most viable candidates are magnetars (\red{neutron stars whose emission is powered by the decay of their magnetic fields}, $B\sim$10$^{13-15}$\,G; \citealt{Bochenek:Ravi:Belov:2020,10.1038/s41586-020-2863-y}), young giant-pulse-emitting pulsars (such as the Crab pulsar; e.g., \citealt{Lyutikov:Burzawa:Popov:2016}), and RRATs \citep[e.g.,][]{Rane:Lorimer:Bates:2015}. However, RRATs and giant pulses pose significant challenges as Galactic counterparts for FRBs. For example, the luminosities of currently known rotationally powered RRATs are at least 8 orders of magnitude lower than those required for typical FRB emission, and FRBs likely require a magnetic, rather than rotational, energy source \citep{Lyutikov:2021}. On the other hand, while giant pulses can approach the luminosity of FRBs, they are often of microsecond duration \citep{Hankins:Kern:Weatherall:2003,Jessner:Popov:Kondratiev:2010}, rather than on the millisecond timescales of the majority of observed FRBs. Hence giant pulses produce much less energy than FRBs.
 
Here we present the discovery of the brightest RRAT observed to date: \pulsar{}, discovered by the Canadian Hydrogen Intensity Mapping Experiment (CHIME). The pulsar can be detected $60^{\circ}$ offset from the meridian of CHIME and peaks at a flux density of $3.1\pm1.0$\,kJy at a dispersion measure (DM) distance of $1.6\pm0.3$\,kpc. Section \ref{sec:obs} details our radio and X-ray observations. We also describe timing and flux calibration methodology. Section \ref{sec:discussion} discusses the implications of our discovery and the potential for finding more such sources.

\section{Observations and Results}
\label{sec:obs}
CHIME is a transit telescope located in British Columbia, Canada. Its wide field of view and large collecting area are particularly well suited to large surveys of radio transients, such as FRBs and pulsars. The CHIME telescope runs several commensal backends, including CHIME \citep{10.3847/1538-4365/ac6fd9}, CHIME/FRB \citep{10.3847/1538-4357/aad188}, CHIME/Pulsar \citep{10.3847/1538-4365/abfdcb}, CHIME/Slow \citep{Mate:Luke:Bhusare:2026}, and the CHIME All-sky Multiday Pulsar Stacking Search (CHAMPSS; \cite{10.3847/1538-4357/adeb51}). The CHIME/FRB experiment has discovered in excess of 100 pulsars thus far\footnote{\url{https://www.chime-frb.ca/galactic}} \citep{Good:Andersen:Chawla:2021,Dong:Crowter:Meyers:2023}. CHIME/FRB uses 1024 digitally formed beams in the CHIME $\sim$2$\times$100 degrees field of view (FOV) to search for Galactic and extragalactic transients.

On 29 September 2025, during a search for bright pulses in CHIME/FRB data, we noted the exceptionally high luminosity of \pulsar{}, whose DM of $\sim$18.5\,pc\,cm$^{-3}$ implies a Galactic origin. This prompted a deep archival search of the sidelobe and main lobe data collected by CHIME/FRB. We also acquired channelized voltage data (baseband data) at CHIME and the three CHIME/FRB outrigger stations \citep{10.3847/1538-4357/adfdcc}. We correlated the CHIME baseband data interferometrically with those obtained at the Outriggers, using the technique detailed in Andrew et al.\ (in prep.). Due to storage constraints, the default data acquisition rules for pulsars at CHIME and the Outriggers record only 100\,ms of data, which limits the number of sufficiently bright in-beam calibrators \citep{Andrew:Leung:Li:2024}. On the longest of the CHIME/FRB outrigger baselines ($\sim$3000\,km), where the calibrator grid is the least dense, we found no in-beam calibrator solution. We were, however, able to obtain in-beam calibrators on the shorter baselines. This resulted in a preliminary localization of RA=21h09m56.93(7)s, Dec=50d55m53(1)s.

Between MJD 59874 and 60955 (22 October 2022 and 7 October 2025) we detected 2275 pulses from \pulsar{}: 1792 in the sidelobes and 471 in the main lobe. \pulsar{} was in the main lobe of CHIME for 574.7\,hours during this period.

The DM measured from \pulsar{} is 18.54(9)\,pc\,cm$^{-3}$. We found no DM evolution over the span of this study.
We infer the distance and the associated uncertainties using the NE2001, YMW16, and NE2025 DM models \citep{Cordes:Lazio:2003,Yao:Manchester:Wang:2017,ocker:cordes:2025}. In these models, the estimated distances are 1.6(3)\,kpc, 1.1(2)\,kpc, and 1.6(3)\,kpc, respectively. Throughout this study, we adopt the NE2025 distance estimate as it uses the most expansive and recent data, including 171 precise pulsar distances based on parallax and globular cluster associations. The median distance prediction accuracy is also 20 and 15 times better than NE2001 and YMW16, respectively \citep{ocker:cordes:2025}.

\subsection{Time of Arrival and Timing}
\label{sec:timing}

\begin{table}[]
\centering
\caption{Properties of \pulsar{}. The first two segments are the timing solution, and subsequently derived quantities from TOAs measured against a standard profile, which was generated from the time-averaged profiles (see Section \ref{sec:timing}). The third segment lists values described throughout the text. Values in parentheses represent the uncertainty on the least significant digit. The burst rates provided are 95\% complete above a flux density of 2.3\,Jy.}
\begin{tabular}{ll}
\hline
\hline
RA ($\alpha$, J2000)          & 21h08m56.751(8)s  \\
Dec ($\delta$, J2000)         & 50d55m53.1(1)s \\
$F$\,(Hz)          & 2.01978077633(3)\\
$\dot{F}$ (Hz/s)   & -2.7345(8) $\times10^{-15}$                                     \\
$P$\,(s)          & 0.49510323681(1)                             \\
$\dot{P}$      & 6.703(2) $\times10^{-16}$                                   \\
EPOCH & 59618\\
DM (pc/cm$^3$) & 18.54(9) \\
\hline
\hline  
$\textrm{T}_{\textrm{age}}$ (yr)      & 1.2 $\times10^{7}$                                  \\
$\textrm{B}_{\textrm{surface}}$ (G)  & 5.8 $\times10^{11}$                                       \\
  $\dot{\textrm{E}}_{\textrm{spindown}}$ (erg/s) & 2.2  $\times10^{32}$                                    \\
\hline
\hline  
Distance NE2001 (kpc) & 1.6(3) \\
Distance NE2025 (kpc) & 1.6(3) \\
Distance YMW16 (kpc) & 1.1(2) \\
Peak Flux Density (kJy) & 3(1)\\
Peak Luminosity (erg\,s$^{-1}$\,Hz$^{-1}$) & 8(2) $\times10^{23}$\\ 
Mode 0 Burst Rate (\#/hr) & 0.31(2)\\ 
Mode 1 Burst Rate (\#/hr) & 0.51(3) \\
Mode 0 Mean Flux (Jy) & 0.035(14) \\
Mode 1 Mean Flux (Jy) & 0.82(32) \\

\hline
\hline
\label{tab:timing_sol}
\end{tabular}
\end{table}
We determine the phase-connected timing solution for \pulsar{} with a two-step process. Initially, for each pulse with total intensity data detected by the CHIME/FRB instrument, the time of arrival (TOA) is defined as the time of the pulse peak. The TOA error is the effective width of the pulse, $W_{\textrm{eff}}=F/S_{\textrm{peak}}$, where $F$ is the fluence, or time-integrated flux density, and $S_{\textrm{peak}}$ is the peak flux density. \red{The fluence is defined as $F=\int_{t_{\rm start}}^{t_{\rm end}}S(t)\,\textrm{d}t$, where $S(t)$ is the median-subtracted flux density time series. $t_{\rm{start/end}}$ are initially defined with a fiducial width of 50\,ms around the pulse peak, which is then refined iteratively by deriving $W$, which in turn puts better estimates on $F$.} We use the effective width as the TOA error because these TOAs serve only to provide an initial period and timing solution. Days with multiple pulses are processed using the \texttt{rrat\_period} program in the \texttt{PRESTO}\footnote{\url{https://github.com/scottransom/presto}}\citep{Ransom:2011} software package to find an initial period of $P\approx$0.495\,s.

To start with, we fix the position of \pulsar{} at the Outrigger-localized position and attempt to phase-connect, allowing only the frequency and its derivative to vary. Due to the presence of two emission modes, which were not known \textit{a priori}, we were unsuccessful in manually phase-connecting the TOAs, so we turned to the Algorithmic Pulsar Timer for Binaries (APTB; \citealt{phillips:ransom:2022,taylor:ransom:padmanabh:2024})\footnote{\url{https://github.com/Jackson-D-Taylor/APT}}. \pulsar{} is the first RRAT to be successfully phase-connected by APTB, marking an expanded use of the algorithm and software that was originally designed for binary millisecond pulsars with sparse data. RRAT TOAs are inherently sporadic, and so tools like APTB may become more essential as CHIME-discovered RRATs increase in number. Due to CHIME's regular sidereal-day cadence, the true spin frequency of a main lobe pulse train can be confused with spin frequencies separated from it by $n/T_{\rm sidereal}$ (where $n$ is an integer), an effect known as aliasing. Sidelobe pulses can be used to establish the correct period, because they can be \red{detected} up to $\sim$4\,hr from the CHIME transit. If the period were aliased, then the minimum timing residual for these pulses would be $\sim0.3$\,s. In other words, an aliased period will cause sidelobe pulses to arrive with seemingly random timing residuals of up to $\pm0.3$\,s for $n=1$, and the residuals will increase for higher $n$. Therefore, to mitigate aliasing, we first phase-connected using only sidelobe TOAs. We then added the main lobe TOAs after determining the correct non-aliased period. More details on aliasing can be found in \citet{10.3847/1538-4357/adeb51} and \citet{Crowter:2025}. After this, we noticed a sinusoid in the residuals of the timing fit due to an incorrect position; therefore, we also fit the position in our initial timing solution.

\begin{figure}
  \centering
  \includegraphics[width=0.49\textwidth]{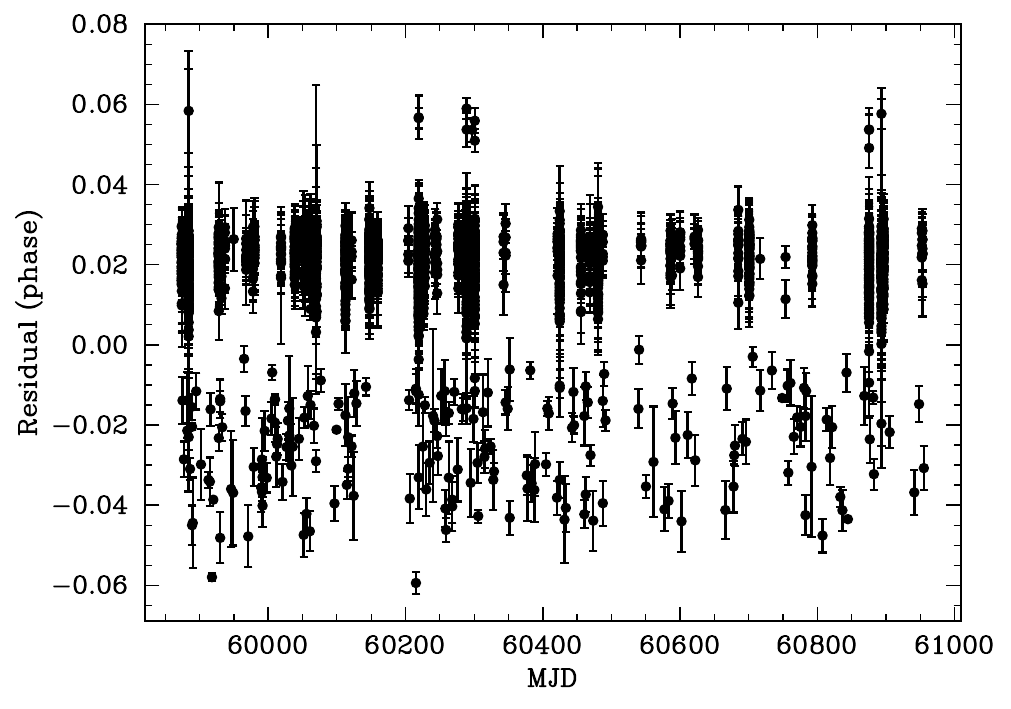}
  \caption{The timing residuals for each pulse detected with total intensity data using CHIME/FRB. The timing residuals are derived with the timing solution presented in Table \ref{tab:timing_sol}.}
  \label{fig:time_residuals}
\end{figure}

Mode changing is commonly observed in pulsars \citep[e.g.,][and references therein]{Wang:Manchester:Johnston:2007}. In RRATs, this is usually between nulling and normal modes \citep[e.g.,][]{Xu:Zhi:Tian:2026}. In addition to RRATs, other pulsars, such as B0823+26, have been reported to exhibit quiet emission modes that are, on average, over 100 times fainter than bright emission modes \citep{Sobey:Young:Hessels:2015}.
With the initial timing solution, we find that \pulsar{} exhibits mode-changing behavior. There are two emission modes with pulses occupying two distinct areas in the timing residual space. The separation in timing residuals can be seen in Figure \ref{fig:time_residuals}. The pulses from each mode arrive with a slight phase offset. One mode is significantly brighter than the other and is discussed in Section \ref{sec:flux_fluence}, with the mean flux differing by a factor of $\sim20$. We label the dimmer mode as mode 0 (negative residuals in Figure \ref{fig:time_residuals}) and the brighter mode as mode 1 (positive residuals in Figure \ref{fig:time_residuals}).

\begin{figure*}
  \centering
  \includegraphics[width=0.8\textwidth]{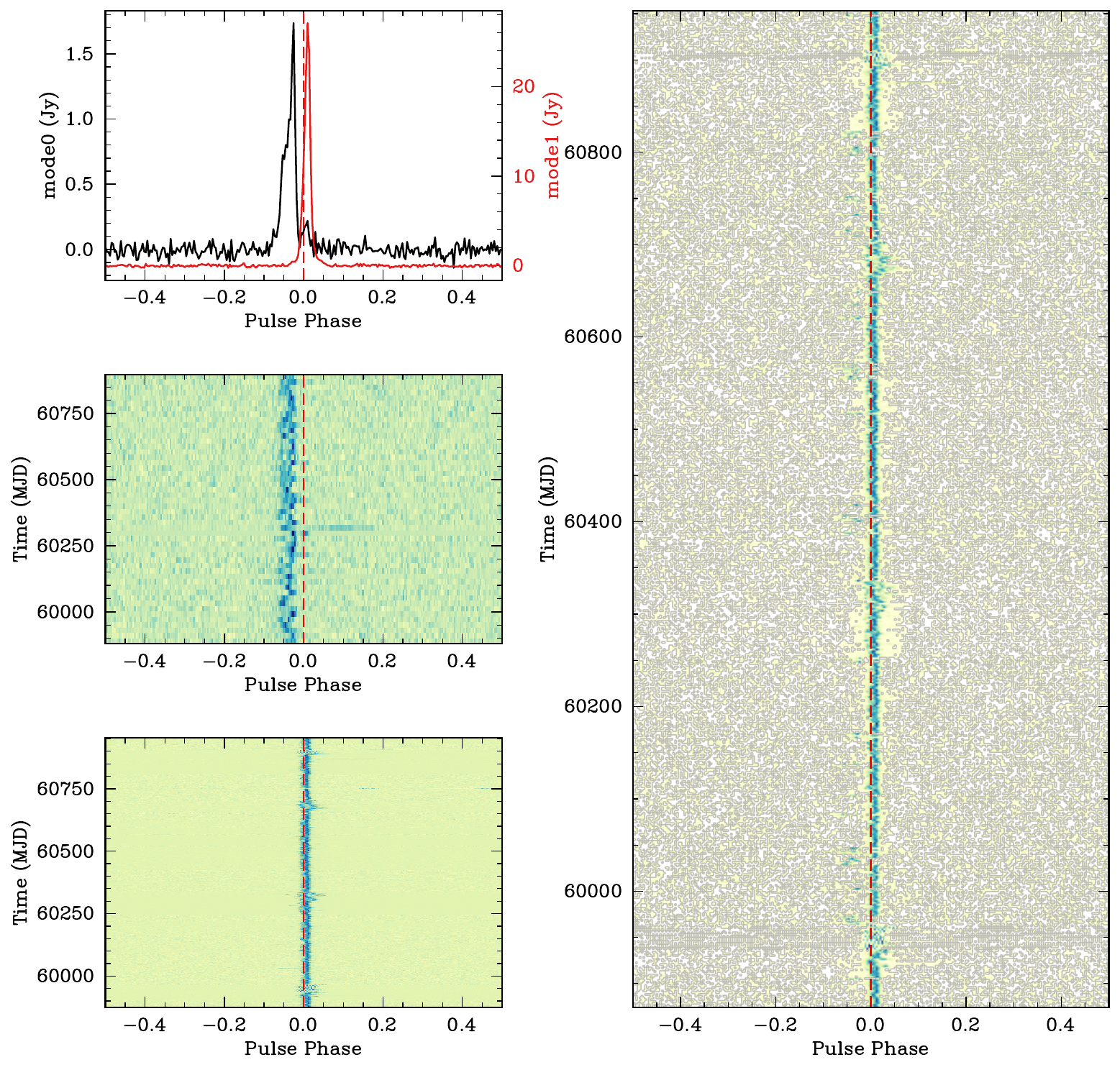}
  \caption{ \red{Left top: The average pulse profile of mode 0 and mode 1. Left middle: The intensity vs phase for mode 0. Left bottom: The intensity vs phase for mode 1. Right: The mean pulse profile of both mode 0 and mode 1. Modes 0 and 1 are independently scaled so that mode 0 is visible; otherwise, mode 1 dominates the color scale. The red dotted vertical line indicates a pulse phase of 0, consistent with other figures in this study.}}
  \label{fig:average_profile}
\end{figure*}

We then refined the timing solution by processing each pulse using the \texttt{dspsr}\footnote{\url{https://sourceforge.net/projects/dspsr/}}\citep{vanStraten:2010} software package to produce \texttt{psrchive} \citep{ 
10.48550/arXiv.1205.6276} files that embed the initial timing solution. This creates a phase-connected \texttt{archive} format file. We separate each emission mode by splitting the data based on the timing residual, creating two pulse profiles. These are shown in Figure \ref{fig:average_profile}. We generate daily \texttt{archive} format files for each emission mode by combining all pulses detected on a given day from a given mode with the \texttt{psradd} program. TOAs are then generated using the daily \texttt{archive} format files and the emission mode-specific pulse profile using the \texttt{pat} program. These TOAs are used to produce the final timing solution using the \texttt{PINT}\footnote{\url{https://github.com/nanograv/PINT}}\citep{Luo:Ransom:Demorest:2021,Susobhanan:Kaplan:2024} pulsar timing package. The final timing solution is presented in Table \ref{tab:timing_sol}.


\subsection{Radio Flux Density Calibration}
\label{sec:flux_fluence}
Radio flux density calibration has long been challenging for the CHIME telescope, with significant uncertainties arising from its transit telescope configuration and cylindrical geometry \citep{Andersen:Patel:Brar:2023}. CHIME calibrates both phase and amplitude gains; therefore, the resulting values after beamforming are already flux-density calibrated relative to the beam model. Hereafter, we refer to this as ``beamformer'' units. For this reason, we avoid the use of a secondary calibrator such as that used by \cite{Andersen:Patel:Brar:2023}, and instead convert directly with the known constants \citep{Andersen:Patel:Brar:2023,Merryfield:Tendulkar:Shin:2022}. Explicitly, for an astrophysical source, the conversion from beamformer units to flux density units is given by
\begin{equation}
  \begin{split}
    S_{src}(\nu,\alpha,\delta,t) =& \frac{1024 f_{good}^{2} \cdot 128}{4^{2}\cdot 0.806745 \cdot 100}\\
                                  &\times\frac{1}{M_{primary}(\nu,\alpha,\delta,t) \cdot M_{formed}(\nu,\alpha,\delta,t)}\\
                                  &\times I_{BF}
    \end{split}
    \label{eq:intensity_fluxcal}
\end{equation}
where $f_{good}$ is the fraction of good feed inputs, $I_{BF}$ is the intensity in beamformer units, and $S_{src}$ is the flux density of the source in Jy. This fraction varies daily, often due to weather conditions, and ranges over 70--95\%\footnote{After 2020 April 23, this correction is no longer needed; instead, it is applied at the gain calibration step.}.

$M_{primary}$ is the primary beam model and $M_{formed}$ is the formed beam model, while $\nu$, $\alpha$, $\delta$, and $t$ are the observing frequency, right ascension, declination, and time of observation, respectively. The formed-beam model is understood analytically and presented in \cite{masui:shaw:ng:2019}.

\begin{figure*}
  \centering
  \includegraphics[width=0.24\textwidth]{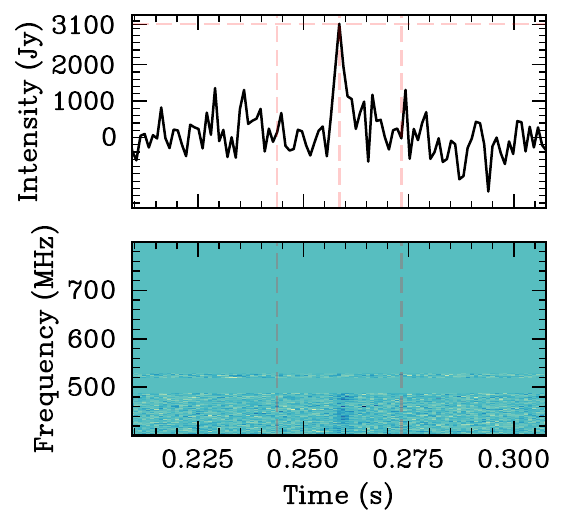}
  \includegraphics[width=0.24\textwidth]{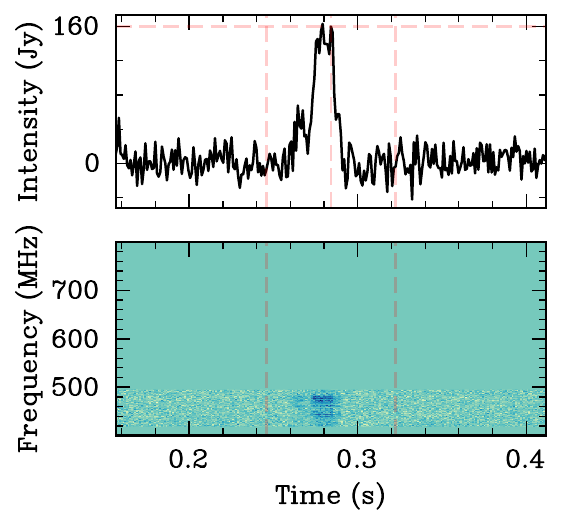}
  \includegraphics[width=0.24\textwidth]{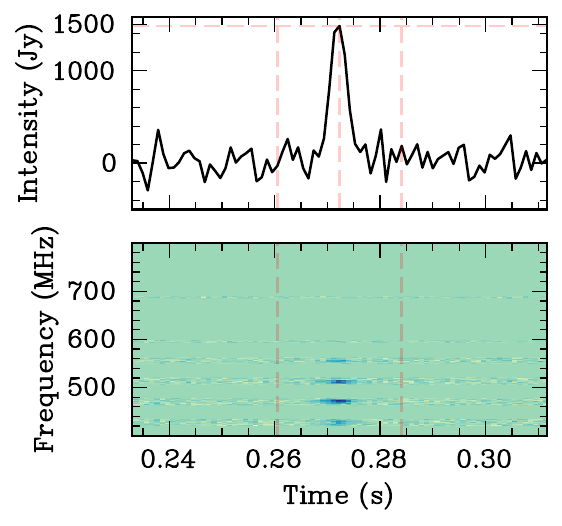}
  \includegraphics[width=0.24\textwidth]{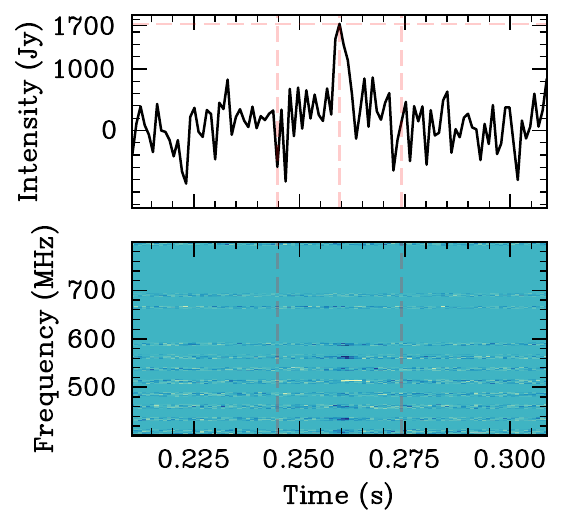}
  \caption{Four notable pulses from \pulsar{}, ordered left to right. \textit{First:} the pulse with the highest peak flux density, $3.1\pm1.0$\,kJy, recorded in a suboptimal formed beam (Appendix \ref{sec:beam_positions}). \textit{Second:} the pulse with the highest fluence. \red{\textit{Third:} the brightest pulse recorded close to its optimal formed-beam response, at $1.5\pm0.7$\,kJy (hour angle $-29.6^\circ$); this is the fiducial pulse used in Section \ref{sec:FRB}. \textit{Fourth:} the brightest pulse detected in the far sidelobes of CHIME (hour angle $-57.7^\circ$). This pulse lies outside the hour angle range covered by our holography data, so its peak flux density of 1.7\,kJy is a lower limit.}  The data shown are the total intensity recorded at 0.98\,ms, with 16,384 frequency channels. The ``clamping'' pattern, characterized by notches in the dynamic spectrum, is a result of \pulsar{} being off-axis \citep{10.23919/URSIGASS.2017.8105318}. In the figure, we apply a static radio frequency interference (RFI) mask to remove known bad frequency channels. We also remove channels with less than 0.1 times the sensitivity of the most sensitive frequency channel. These channels usually contain no signal and thus only dilute the pulse. The pink lines show the maximum flux density and the integration region for the fluence calculation.}
  \label{fig:brightest_burst}
\end{figure*}

Due to the high flux densities of \pulsar{}, we detect a large number of pulses in the sidelobes. CHIME has larger sidelobes than telescopes such as the Green Bank Telescope (GBT) due to its cylindrical aperture and the relatively smaller extent of each cylinder (20\,m, compared to the 100\,m of the GBT). Sensitivity in the CHIME sidelobes is attenuated by factors of only $10^{-2}-10^{-3}$ relative to the main lobe \citep{reda:Pinsonneault-Marotte:Deng:2022,lin:scholz:ng:2024}, which, for a source as bright as \pulsar{}, still permits detection. This results in an effective FOV of 24.44$^\circ\times$100$^\circ$ (R.A.$\times$Dec) with an attenuation above $10^{-2}$. In our study, we differentiate between sidelobe and main lobe pulses using a fiducial cutoff of 2$^{\circ}$ in hour angle from the meridian. \red{In Figure \ref{fig:brightest_burst} we show four notable pulses from \pulsar{}: the pulse with the highest peak flux density, the pulse with the highest fluence, the brightest pulse recorded close to its optimal formed-beam response, and the brightest pulse detected in the far sidelobes of CHIME. The pulse with the highest peak flux density was detected in a near sidelobe, but only data from a sub-optimally formed beam in CHIME/FRB's beam grid were saved, where the beam response is smaller and varies more rapidly across the band. As a result, we caution that the flux-density calibration for this burst carries high uncertainty; the formed beam response for this pulse is shown in Appendix \ref{sec:beam_positions}. The brightest far-sidelobe pulse was detected at an hour angle of $-57.7^\circ$, outside the range covered by our holography data (Section \ref{sec:flux_fluence}), so its peak flux density of 1.7\,kJy is a lower limit.}

\begin{figure}
  \centering
  \includegraphics[width=0.45\textwidth]{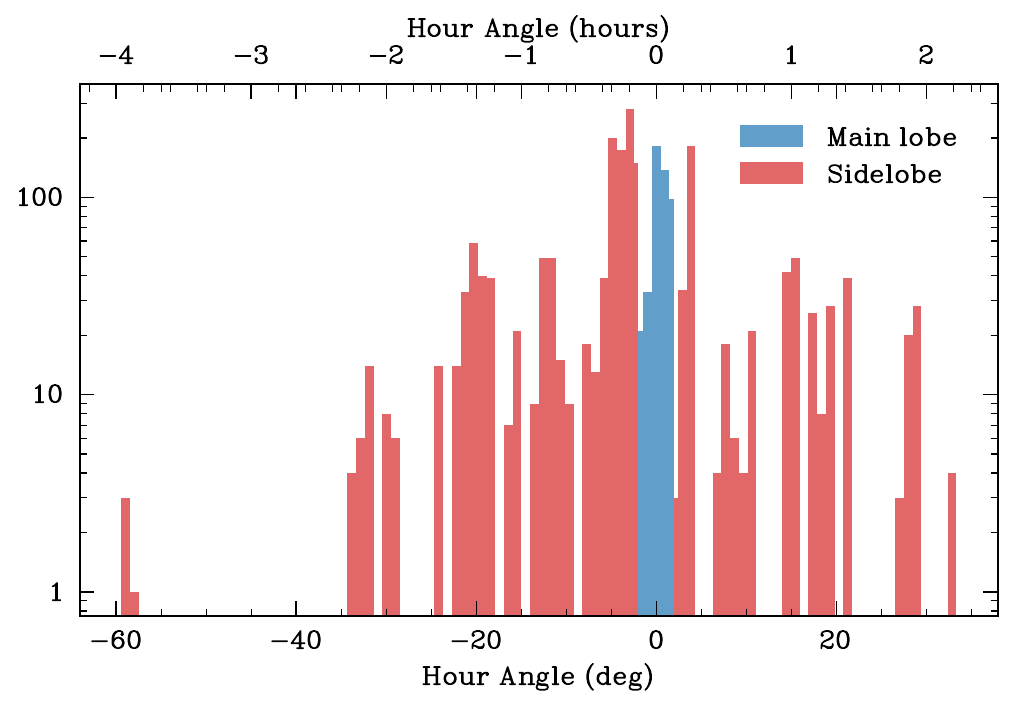}
  \caption{The hour angle distribution of pulses from \pulsar{}. The sidelobes are defined as $>2^\circ$ hour angle from the CHIME transit.}
  \label{fig:ha_hist}
\end{figure}

\begin{figure}
  \centering
  \includegraphics[width=0.49\textwidth]{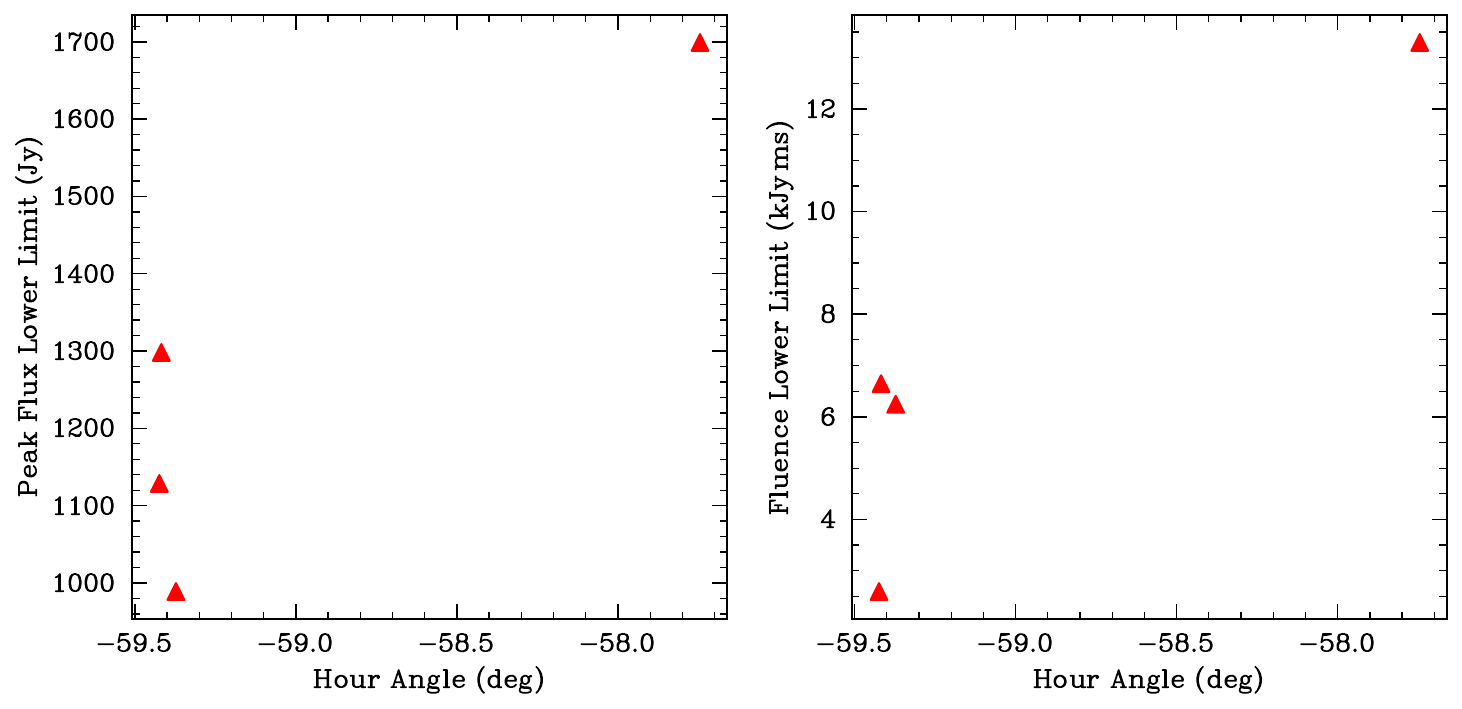}
  \caption{Radio pulses that were detected outside of the hour angle range for which we have reliable holography data. The flux density and fluence values for these pulses are therefore lower limits.}
  \label{fig:flux_lower}
\end{figure}

The standard beam model for CHIME is reliable only for sources within the main lobe \red{\citep{10.3847/1538-4365/ac6fd9}}, not in the sidelobes. Hence, to create the primary-beam model for the sidelobes, we employ holography between the John A. Galt Telescope (Galt Telescope) and each CHIME feed. This technique has already been shown to be effective for characterizing the CHIME far sidelobes and for correcting for the brightness of FRBs in the far sidelobes \citep{reda:Pinsonneault-Marotte:Deng:2022,lin:scholz:ng:2024}. The basic principle is that the 26\,m Galt telescope tracks a bright source as it moves across the CHIME sky. The recorded signal from the Galt telescope is then cross-correlated with each CHIME feed, yielding a CHIME response to that source. The benefit of such an approach is that, as the signal weakens due to beam-response attenuation, the astrophysical source can still be retrieved from the correlation process, rather than beamforming towards the bright source with CHIME alone. In other words, holography reduces the confusion noise on a given point source. In this study, we use holography tracks of Cygnus A to measure the beam response in the far-sidelobe regime. A detailed explanation of the procedure is given by Dong et al.\ (in prep.). In Appendix \ref{sec:holography} we present the resultant beam response and validation flux density estimates for Cygnus A and Taurus A in the sidelobes. We use only holographic data for Cygnus A, as it is the closest holography track to the declination of \pulsar{}. The longest holography track that we have of Cygnus A is 6 hours, centered on the transit. The distribution of pulses in relation to the CHIME transit is shown in Figure \ref{fig:ha_hist}. In addition to the pulses we were able to flux-calibrate, we detected four pulses outside the range covered by our holography data; for these, we provide only lower limits on flux density and fluence. A summary of these pulses is provided in Figure \ref{fig:flux_lower}. \red{The brightest of these, detected at an hour angle of $-57.7^\circ$ and shown in the rightmost panel of Figure \ref{fig:brightest_burst}, has a lower limit on its peak flux density of 1.7\,kJy, and its intrinsic flux density could be substantially higher.} All measured pulse parameters are provided along with this publication.

\begin{figure*}
  \centering
  \includegraphics[width=0.7\textwidth]{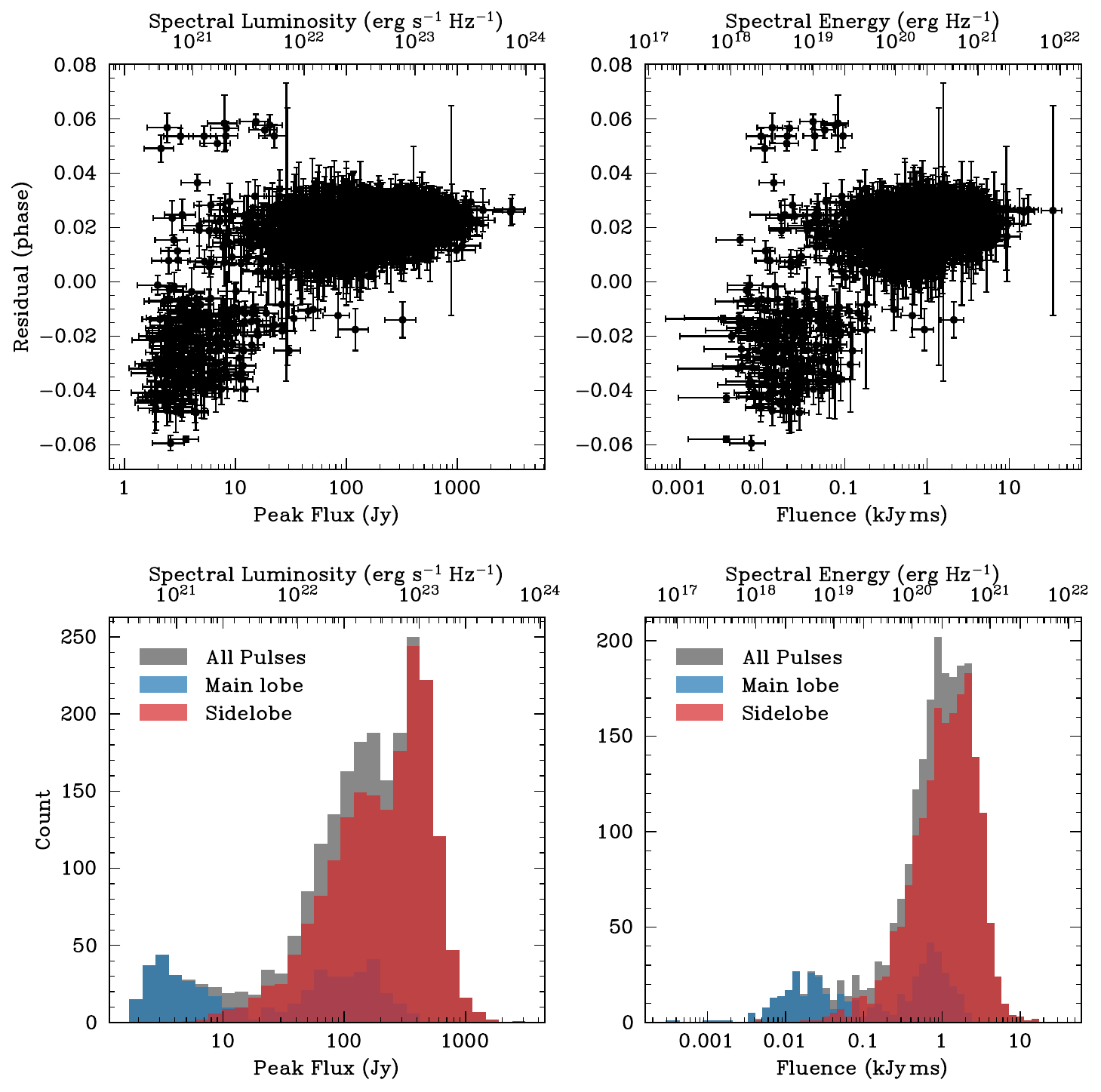}
  \caption{The top row shows the timing residuals plotted against the peak flux density and fluence. We note that across both timing residual figures, there is a clear separation between two distinct emission modes, as indicated by both intrinsic brightness and residuals. The bottom row shows histograms of the peak flux and fluence. The histograms appear bimodal; however, these histograms are not corrected for the complex selection effects of the CHIME/FRB system.}
  \label{fig:resid_flux_fluence}
\end{figure*}

We compiled the peak flux density and fluence distribution of all pulses detected by CHIME/FRB. These are presented in Figure \ref{fig:resid_flux_fluence}. The brightest peak flux density was $S_{\textrm{peak}}=3.1\pm1.0$\,kJy, and the highest-fluence pulse had $F=34\pm9$\,kJy\,ms at 600\,MHz. These correspond to a peak spectral luminosity of $(7.5\pm2.3)\times10^{23}$\,erg\,s$^{-1}$\,Hz$^{-1}$ and a spectral energy of $(8.2\pm2.2)\times 10^{21}$\,erg\,Hz$^{-1}$ at the assumed DM-derived distance of 1.6\,kpc. Both quantities follow the convention $L_\nu = d^2 S_\nu$ (i.e., an emitting solid angle of 1\,sr), as adopted in the transient phase space literature \citep{Nimmo:Hessels:Kirsten:2022}. \red{The brightest pulse recorded close to its optimal formed-beam response, at an hour angle of $-29.6^\circ$, has $S_{\textrm{peak}}=1.5\pm0.7$\,kJy, corresponding to a peak spectral luminosity of $(3.6\pm1.6)\times10^{23}$\,erg\,s$^{-1}$\,Hz$^{-1}$. We note that the brightest far-sidelobe pulse, at 1.7\,kJy, is brighter still, but falls outside the hour angle range of our holography data and so is only a lower limit (Figure \ref{fig:flux_lower}).} The brightest pulses are all emitted in mode 1, and in this mode they appear nearly persistent across the activity window (Section \ref{sec:champss}). To the best of our knowledge, among single pulses from pulsars, only giant pulses from PSR B0531+21 (the Crab pulsar; \citealt{Cordes:Bhat:Hanskins:2004,Jessner:Popov:Kondratiev:2010,Sokolowski:Kumar:Dhavali:2025}) are more luminous than those from \pulsar{}. We note that PSR B0329+54 may produce pulses of similar peak brightnesses, however, we did not find any direct evidence in the literature. We also note that magnetars, such as SGR 1935+2154, and FRBs can also produce single bursts brighter than \pulsar{} \citep{Bochenek:Ravi:Belov:2020,10.1038/s41586-020-2863-y}.



\subsection{CHAMPSS Detection}
\label{sec:champss}
\begin{figure}
  \centering
  \includegraphics[width=0.49\textwidth]{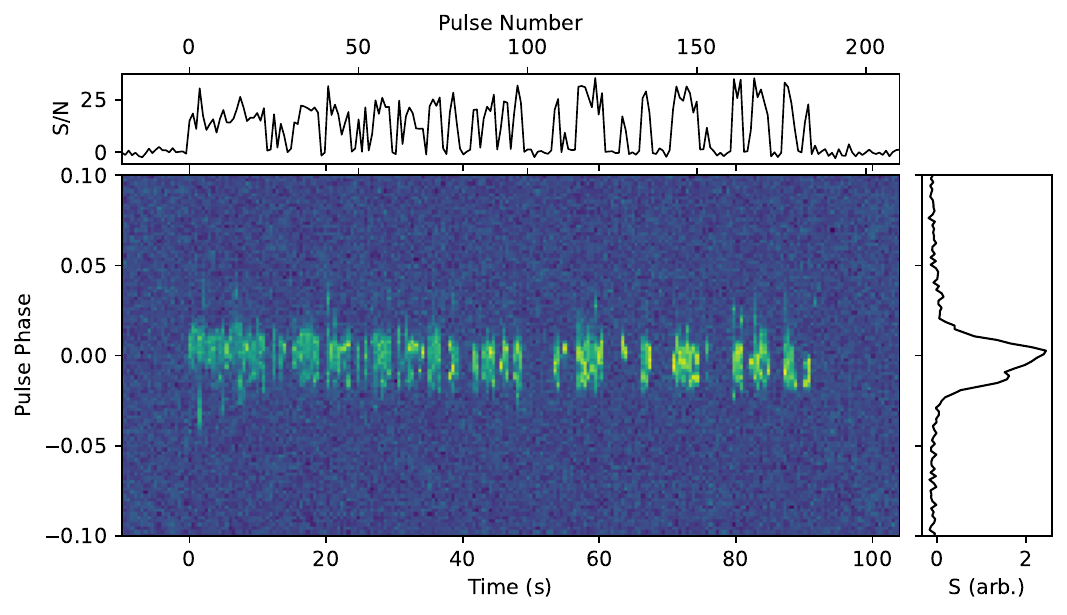}
  \caption{Activation of \pulsar{} detected with the CHAMPSS system.}
  \label{fig:CHAMPSS}
\end{figure}

In addition to the CHIME/FRB single pulse detections presented thus far, \red{both modes of} \pulsar{} was blindly detected as a periodic source by the CHAMPSS system \citep{10.3847/1538-4357/adeb51}, which performs daily and stacked periodicity searches using CHIME/FRB data.
A declination range of $49^{\circ}$--$61^{\circ}$ was observed daily by CHAMPSS from MJD 61046 to 61076 (2026 January 6 to 2026 February 5). On MJD 61065, a 47.8-$\sigma$ candidate was detected with DM=18.55\,pc\,cm$^{-3}$ and a spin frequency of 2.01963\,Hz. The source was identified as \pulsar{} in a near sidelobe, with the detection spanning 1.8$^\circ$--1.3$^\circ$ before culmination. When the source is on, it appears to be a persistent periodic source; the activity began and ended suddenly, with pulses detected in the majority of rotations in this state. All emission components shown in Figure \ref{fig:resid_flux_fluence} are present, sometimes occurring in the same rotation. This is shown in Figure \ref{fig:CHAMPSS}.
This $\sim 90$-second ``on'' mode was the only detection of \pulsar{} through the full month of daily observations. The full transit length corresponds to $\sim$1600\,s, from which we can infer a duty cycle of the bright mode of $\lesssim 0.002$. 

\subsection{High Energy Observations}
Given the source's remarkable radio luminosity properties, we investigated whether it is also an unusual high-energy emitter.
We observed the position of \pulsar{} with the {\it Neil Gehrels Swift Observatory} X-ray Telescope for 781\,s (observation ID: 3000165) and found no sources within the localization region of \pulsar{}. We analyzed the data using the University of Leicester XRT data analysis portal \footnote{\url{https://www.swift.ac.uk/user_objects/}}. The background rate for this pointing was $9.8\times10^{-3}$\,counts\,s$^{-1}$ between 0.2--10\,keV. Taking a Galactic $N_\mathrm{H}=1.1\times 10^{22}$\,cm$^{-2}$ \citep{Dickey:Lockman:1990}, we estimate an X-ray flux upper limit of $2\times 10^{-13}$\,erg\,cm$^{-2}$\,s$^{-1}$ and an unabsorbed flux of $7.7\times 10^{-13}$\,erg\,cm$^{-2}$\,s$^{-1}$, using a blackbody spectrum of kT=0.13\,keV \citep{10.1086/521387} between 0.2--10\,keV. From the DM-derived distance of 1.6\,kpc we determine an X-ray luminosity upper limit of $6\times 10^{31}$\,erg\,s$^{-1}$ between 0.2--10\,keV. This limit is 5 orders of magnitude lower than the X-ray luminosity of the Crab pulsar \citep{10.3847/1538-4357/aad911}.
\subsection{Radio Continuum and Optical}
\red{We further searched for counterparts at other wavelengths. In particular, we searched radio continuum imaging from the Very Large Array Sky Survey (VLASS) and the Panoramic Survey Telescope and Rapid Response System (Pan-STARRS) imaging archive in the $g$, $r$, $i$, $z$, and $y$ bands. We find no counterpart to \pulsar{} within $3\sigma$ of the timing position.}
\section{Discussion}
\label{sec:discussion}

\subsection{Quantifying the Number of Undiscovered Extremely-luminous RRATs}
\label{sec:GBNCC}
The discovery of \pulsar{} is surprising in that numerous other surveys, such as the Green Bank North Celestial Cap survey (GBNCC; \citealt{Stovall:Lynch:Ransom:2014}) and the High Time Resolution Universe North survey (HTRU North; \citealt{Barr:Champion:Kramer:2013}), have missed it. This section explores the underlying population of extremely luminous RRATs.

\begin{figure}
  \centering
  \includegraphics[width=0.48\textwidth]{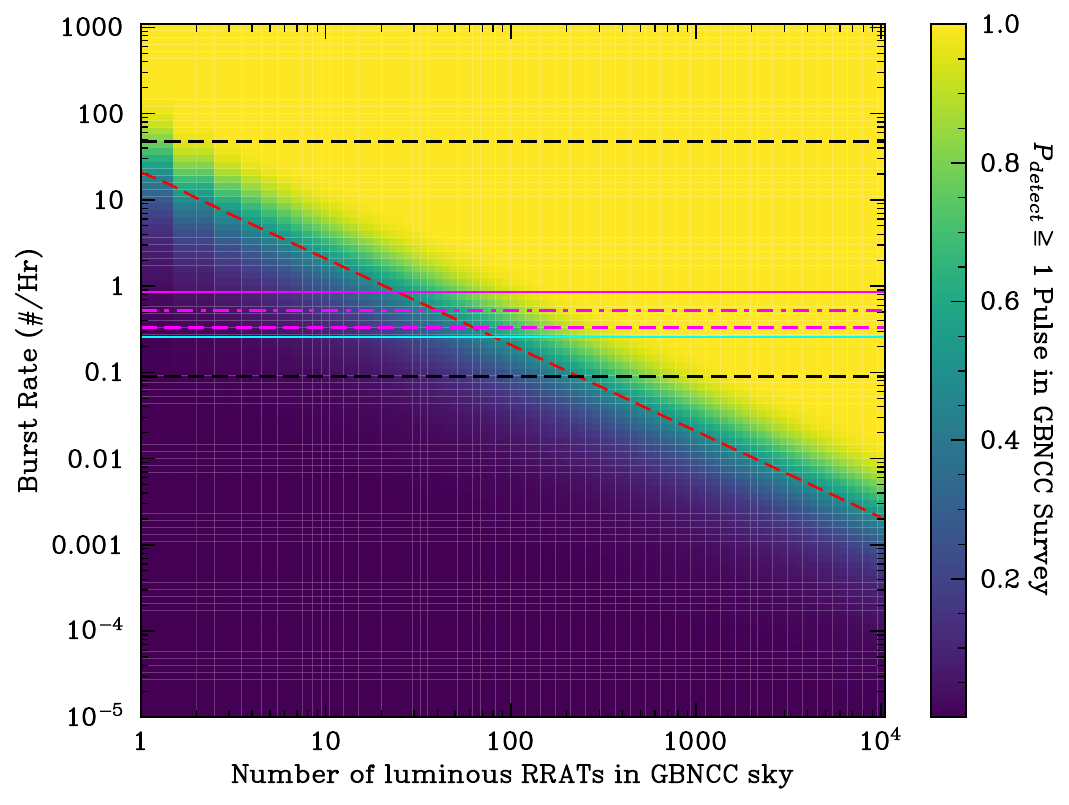}
  \caption{The probability of detecting at least one pulse from one extremely luminous RRAT in the GBNCC Survey. The black-dashed line shows the mean and minimum burst rates of RRATs in the RRATalogue. We note that these rates are likely biased high, as RRATs with low burst rates have no measurement. The red dashed diagonal line is the $P=0.5$ contour. The magenta solid, dash dot, and dashed lines (in descending order) show the burst rate of \pulsar{} in aggregate, mode 1, and mode 0, respectively. The cyan solid line shows the $S_{peak}>100$\,Jy \pulsar{} burst rate, where $S_{peak}$ is the peak flux density. The \pulsar{} burst rates are provided in Table \ref{tab:timing_sol}. }
  
  \label{fig:GBNCC_toy}
\end{figure}

\pulsar{} may be the tip of the iceberg for highly luminous sporadic pulsars missed by previous surveys. Because our survey has complex selection effects and the data have not been fully searched, it is difficult to estimate the population of \pulsar{}-like sources present in our data. However, we use a toy model to estimate the number of bright RRATs in the GBNCC sky. In this calculation, we are interested only in RRATs with $S_{peak}>100$\,Jy. For GBNCC, the minimum sensitivity for single pulses is $\sim0.25-10$\,Jy for DMs between $0-1000$\,pc\,cm$^{-3}$ \citep{Karako-Argaman:Kaspi:Lynch:2015}. Therefore, in our model, we assume that if such an RRAT lies within a GBNCC pointing and emits a pulse with peak flux density greater than 100\,Jy during that pointing, the detection probability is 100\%. Next, we generate a grid of the number of bright RRATs, $N$, and burst rates, $B$. As GBNCC covered the entire northern sky, we can assume that all bright RRATs received the full $T=120$\,s dwell time \citep{Stovall:Lynch:Ransom:2014}. The probability of detecting at least one pulse from at least one extremely luminous RRAT is then
\begin{equation}
  P_{\geq 1}= 1-\prod_{i=1}^{N}P_{i,0},
\end{equation}
where $P_{i,0}$ is the probability of detecting no pulses from RRAT $i$. Assuming Poissonian burst statistics,
\begin{equation}
    P_{i,0}=e^{-BT}.
\end{equation}
We show the full contour of the probability of detection in Figure \ref{fig:GBNCC_toy}. \pulsar{} emits $S_{\textrm{peak}}>100$\,Jy pulses at a rate of 0.22\,hr$^{-1}$. At this rate, GBNCC would have had a $>50$\% chance of detecting at least one such source if more than 82 had been present, so we take $N\lesssim82$ as our upper limit on the number of \pulsar{}-like sources in the GBNCC sky. We note that, in this toy model, we assume all extremely-luminous RRATs have the same burst rate and flux distribution, since \pulsar{} is the first one discovered.

In Figure \ref{fig:GBNCC_toy} we show the mean and minimum burst rates derived from the RRATalog\footnote{\url{https://github.com/rratalog/rratalog}}\citep{Agarwal:Lewis:Lorimer:2026}. However, these values should be treated as upper limits on the typical burst rate, since RRATs with burst rates too low to measure are absent from the catalog and so do not pull the mean down. Furthermore, long-dwell-time surveys such as CHIME/FRB have systematically uncovered RRATs with higher nulling fractions than those in previous surveys \citep[e.g.][]{Dong:Crowter:Meyers:2023}. This indicates that surveys like GBNCC have only probed the population of high-burst-rate RRATs. For this reason, a population of very luminous RRATs likely remains undiscovered.

A few other surveys are forthcoming that will fill this long-dwell-time niche. Specifically, the Bustling Universe Radio Survey Telescope in Taiwan \citep[BURSTT;][]{Hsiu-Hsien:Kai-yang:Chao-Te:2022}, and the Coherent All Sky Monitor \citep[CASM;][]{Connor:Ravi:Sanghavi:2026} are wide-field-of-view, long-dwell-time surveys with system equivalent flux densities of $\sim$5000\,Jy and $\sim$2750\,Jy, respectively. These are ideal instruments with which to conduct such a survey. Pairing them with a high-sensitivity instrument such as the Green Bank Telescope, which could follow up their candidates, would unambiguously identify RRATs that emit high-luminosity pulses. Furthermore, the CHIME/FRB and CHAMPSS surveys will continue to monitor the sky for sporadic pulsars.

\subsection{Implications for Fast Radio Bursts}
\label{sec:FRB}
\begin{figure*}
  \centering
  \includegraphics[width=\textwidth]{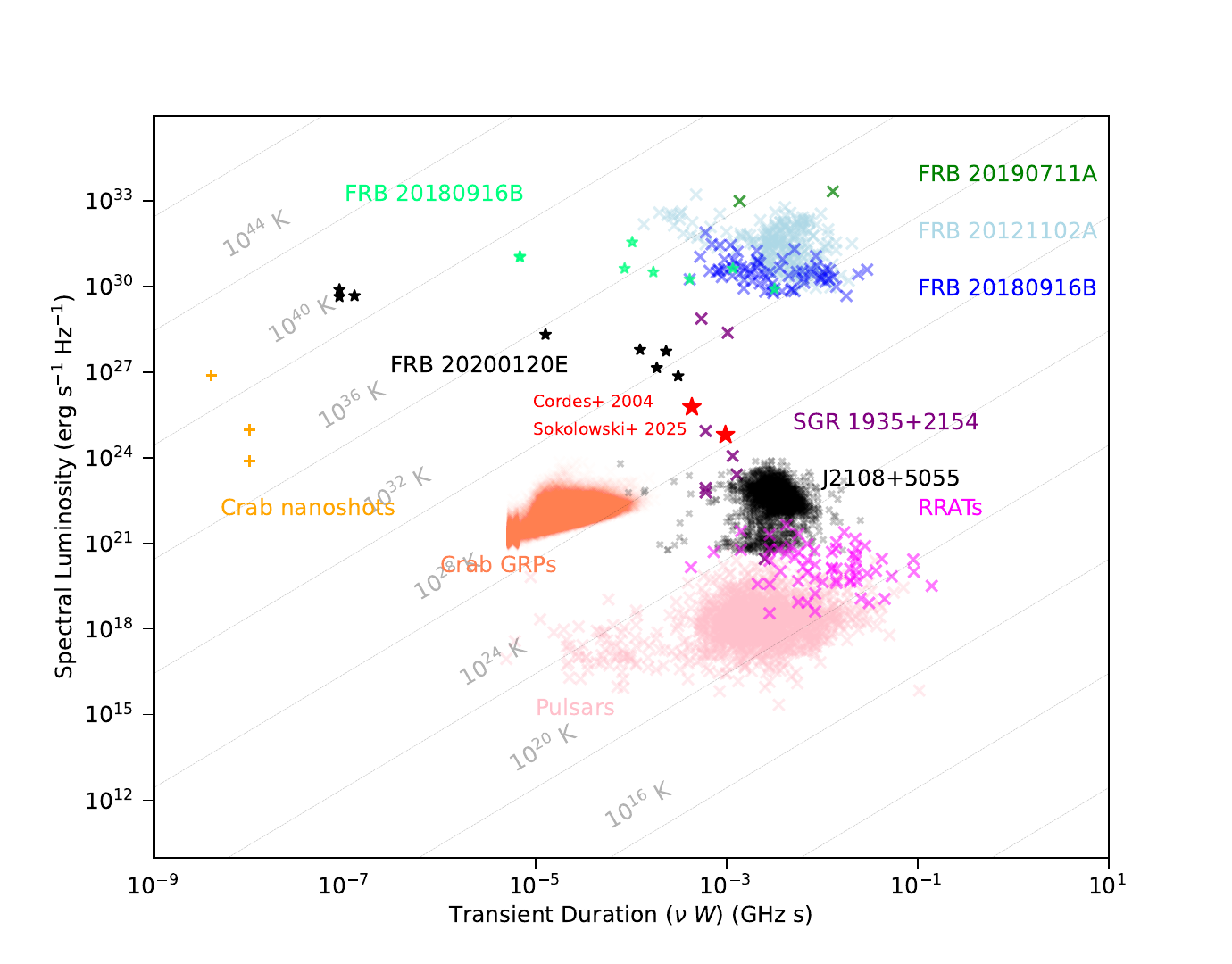}
  \caption{Transient phase space figure for \pulsar{} in comparison to other transients. \pulsar{} reaches a spectral luminosity similar to that of intermediate SGR1935+2154 bursts. Figure adapted from \cite{Nimmo:Hessels:Kirsten:2022}. Sokolowski+ (\cite{Sokolowski:Kumar:Dhavali:2025}) represents the brightest detected Crab giant pulse at 215\,MHz. Cordes+ (\cite{Cordes:Bhat:Hanskins:2004}) represents the brightest Crab giant pulses detected at 430\,MHz. The widths of Crab giant pulses at low frequencies are inflated due to scattering.}
  \label{fig:TPS}
\end{figure*}
Most current evidence suggests that the central engines of FRBs are neutron stars; for example, they could take the form of magnetar bursts \citep[e.g.,][]{Bochenek:Ravi:Belov:2020,10.1038/s41586-020-2863-y}. Some have suggested that FRBs must be powered by magnetic energy rather than rotational kinetic energy \citep{Lyutikov:2021}. In addition, it has been shown that giant pulses from the Crab pulsar can reach flux densities of 16\,kJy at 215\,MHz \citep{Sokolowski:Kumar:Dhavali:2025}, and $\sim155$\,kJy at 430\,MHz \citep{Cordes:Bhat:Hanskins:2004}. These Crab giant pulses could be detected from nearby galaxies \citep{Mclaughlin:Cordes:2003}. The giant pulses are known to follow a power-law distribution, in contrast to those from normal pulsars, which follow log-normal distributions \citep{Lundgren:Cordes:Ulmer:1995,Mickaliger:McEwen:McLaughlin:2018,Burke-Spolaor:2012}. To date, no high-energy cutoff for giant pulses has been observed, suggesting the potential to emit at much higher luminosities. A key challenge remains, however, for explaining FRB emission with Crab-like central engines: Crab giant pulses last only microseconds \citep{Hankins:Kern:Weatherall:2003,Jessner:Popov:Kondratiev:2010}, whereas the majority of FRBs are of millisecond duration \citep{10.3847/1538-4365/ac33ab}, even though some microsecond-duration bursts have been seen \citep{Nimmo:Hessels:Kirsten:2022,Snelders:Nimmo:Hessels:2023}. This mismatch may be exaggerated by selection effects in the CHIME/FRB experiment, given its minimum search time resolution of 0.98\,ms, but it is unlikely that giant pulses alone can account for the observed FRB durations. \pulsar{}, by contrast, emits on a millisecond timescale, like many observed FRBs, although its luminosities remain well below those of the bulk of the FRB population. We stress that we are not proposing \pulsar{} as an FRB progenitor: at $\sim10^{24}$\,erg\,s$^{-1}$\,Hz$^{-1}$ it still falls some eight orders of magnitude short of a typical FRB.
In Figure \ref{fig:TPS}, we place \pulsar{} in context with Crab giant pulses, magnetar bursts, and FRBs. 

In the following, we estimate the S/N of \pulsar{} if it were placed in a nearby galaxy. We use the Five-Hundred-Meter Aperture Spherical Telescope (FAST) due to its high sensitivity. Using a system temperature of $T_{sys}=24$\,K, a gain of $G=16$\,K/Jy, a bandwidth of $\delta \nu = 400$\,MHz \citep{Jiang:Tang:Hou:2020}, a pulse width of $\delta t=10$\,ms, the number of polarizations $n_p=2$, and a Gaussian correction factor of $\eta=0.868$ we apply the single pulse radiometer equation
\begin{equation}
 S/N = S_{peak}\frac{\eta G\sqrt{n_p \delta \nu \delta t}}{T_{sys}}.
\end{equation}
We scale from two reference pulses, both shown in Figure \ref{fig:brightest_burst}. \red{As our fiducial case, we adopt the brightest pulse recorded close to its optimal beam shape, at $1.5\pm0.7$\,kJy (third panel). We also quote the brightest pulse detected from \pulsar{}, at $3.1\pm1.0$\,kJy (leftmost panel), as an optimistic case, bearing in mind its noisier calibration (Section \ref{sec:flux_fluence}). We do not scale from the brightest far-sidelobe pulse (rightmost panel), whose 1.7\,kJy peak flux density is only a lower limit, but we note that it would place the fiducial reach somewhat further out.}

A \pulsar{}-like pulsar in the Andromeda Galaxy (M31, at 765\,kpc; \citealt{Riess:Fliri:Vallas-Gabaud:2012}) would be detected with S/N=22.2 for the 3.1\,kJy pulse and \red{S/N=10.7 for the 1.5\,kJy pulse. At 3.1\,kJy, the most distant galaxy in which FAST would recover such pulses above S/N=6 is Sextans B, at the edge of the Local Group (1.37\,Mpc; \citealt{10.1086/382905}), with S/N=6.9. At the conservative 1.5\,kJy, the reach shrinks to $\sim1$\,Mpc, reaching the Aquarius Dwarf (988\,kpc) at S/N=6.4, while the same pulse in Sextans B would fall to S/N=3.3.} Either way, \pulsar{}-like pulses are detectable throughout the inner Local Group. These estimates neglect any additional scattering or dispersion smearing incurred in the host galaxy and the intergalactic medium, which would broaden the pulse and reduce the recovered S/N. Conversely, we have no evidence that the pulses reported here represent the highest luminosity \pulsar{} can reach.


\begingroup\color{black}
\subsection{Radio efficiency and the Spindown Energy Budget}
\label{sec:efficiency}
From the timing solution, the spindown luminosity of \pulsar{} is $\dot{E}=2.2\times10^{32}$\,erg\,s$^{-1}$. Assuming that the beam remains steady in between pulses, the radio power emitted while the pulsar is in its bright mode is steady and can be compared directly against $\dot{E}$, 
\begin{equation}
  L_{\rm radio} = \Omega_{\rm b} d^2 S_{\rm peak} \Delta\nu,
\end{equation}
where $\Omega_{\rm b}$ is the solid angle of the emission beam, $d=1.6$\,kpc is the DM-derived distance, $S_{\rm peak}=3.1$\,kJy is the peak flux density of the brightest pulse, and $\Delta\nu=400$\,MHz is the CHIME bandwidth.

We take $\Omega_{\rm b}=\pi\rho^2$ from the beam radius $\rho=4.8^\circ(1+66/\nu_{\rm MHz})P^{-0.5}$ of \citet{Mitra:Deshpande:1999}, where $\nu_{\rm MHz}$ is the observing frequency in MHz; at 600\,MHz and $P=495$\,ms this gives $\rho=7.6^\circ$ and $\Omega_{\rm b}=0.055$\,sr. We do not use the $L_\nu=d^2S_\nu$ convention adopted for the luminosities quoted elsewhere in this work, which is a convention rather than a reference value of $\Omega_{\rm b}$. The radio efficiency is then
\begin{equation}
  \eta = \frac{L_{\rm radio}}{\dot{E}} = 7.5\pm3.7\,\%,
\end{equation}
where the uncertainty is dominated by the DM distance. This assumes the beam is uniformly as bright as it appears when its centre crosses our line of sight, which for a source as variable as \pulsar{} is uncertain.

Several percent is high for a rotation-powered pulsar, which typically converts $10^{-6}$--$10^{-4}$ of its spindown power into radio emission, but the figure applies only to the brightest rotations, which are rare. \pulsar{} has a duty cycle of $\lesssim0.002$ (Section \ref{sec:champss}), and the majority of rotations produce no detectable pulse at all. Averaged over time, the radio output is therefore $\sim10^{-4}$ of $\dot{E}$, at the upper end of the range for ordinary rotation-powered pulsars.
\endgroup

\section{Conclusion}
In conclusion, we have discovered an extremely luminous but sporadic mode-changing rotating radio transient. We show that pulses from \pulsar{} can reach $3.1\pm1.0$\,kJy, corresponding to a spectral luminosity of $(7.5\pm2.3)\times 10^{23}$\,erg\,s$^{-1}$\,Hz$^{-1}$. \red{The brightest pulse recorded close to its optimal formed-beam response reaches $1.5\pm0.7$\,kJy, or $(3.6\pm1.6)\times10^{23}$\,erg\,s$^{-1}$\,Hz$^{-1}$, and a further pulse detected in the far sidelobes exceeds 1.7\,kJy, a lower limit set by the hour angle coverage of our holography data.} Using a simple model, we demonstrate that conventional surveys such as GBNCC are likely to miss such sources due to their short dwell times, and that a population of highly luminous RRATs therefore likely remains undiscovered; how large that population is will depend on the intermittency of these sources. We expect CHIME/FRB and CHAMPSS to uncover more of them in the future. Finally, we show that a \pulsar{}-like source placed anywhere in the inner Local Group would be detectable by FAST as extragalactic bursts.
\begin{acknowledgments}
\allacks{}
\end{acknowledgments}

\begin{contribution}
\end{contribution}
\facilities{CHIME, Swift(XRT)}

\clearpage
\appendix

\section{Holography}
\label{sec:holography}
\begin{figure}
  \centering
  \includegraphics[width=0.49\textwidth]{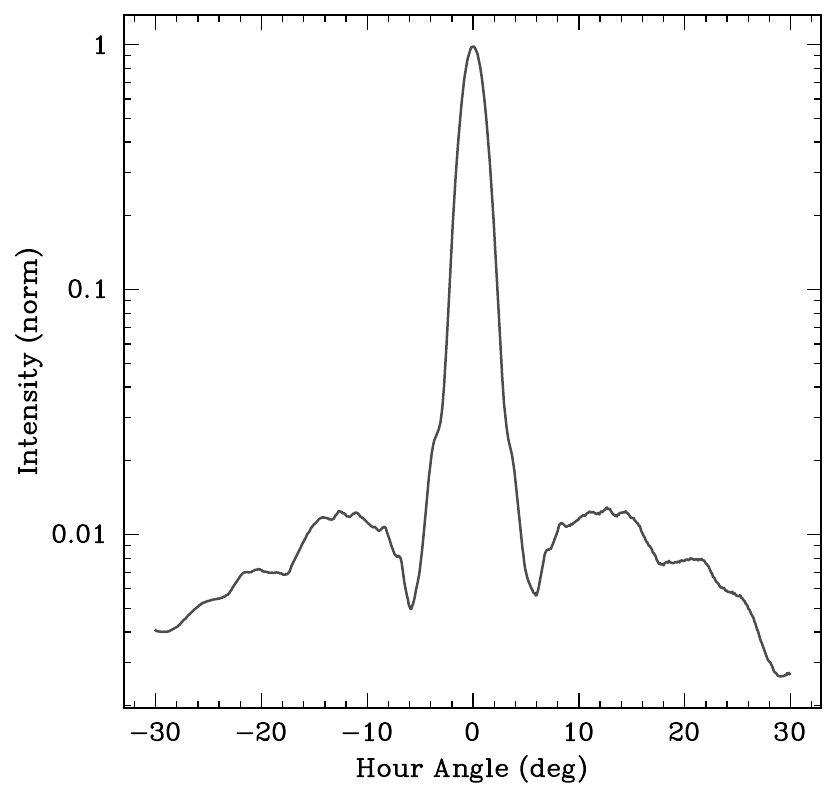}
  \includegraphics[width=0.49\textwidth]{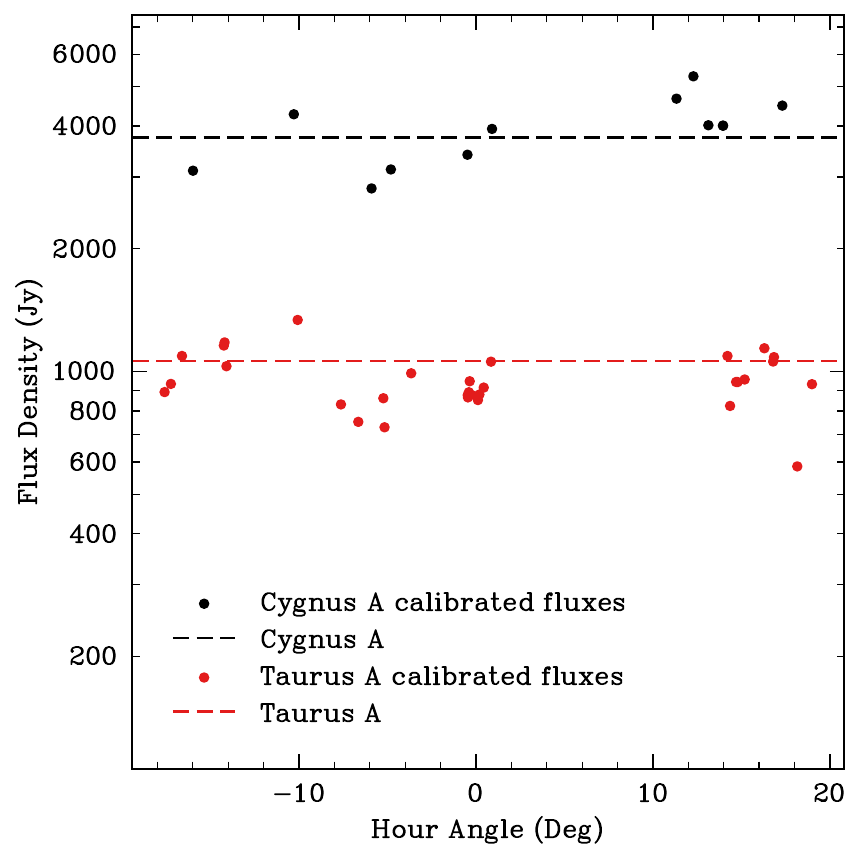}
  \caption{Left: Band-averaged CHIME response normalized to the sensitivity of CHIME towards Cygnus A at the meridian. These data were derived from holography between the Galt Telescope and individual CHIME feeds. Right: Calibration of the flux density of Cygnus A and Taurus A in the sidelobes.}
  \label{fig:holography}
\end{figure}
The beam response of CHIME normalized to Cygnus A is given in Figure \ref{fig:holography}. This shows the mean beam response across the whole CHIME 400--800\,MHz band. The uncertainties on our flux densities are derived in \cite{Dong:inprep}, and we summarize them here. Using this beam response and raw channelized voltage data beamformed on and off Cygnus A and Taurus A, both sources are recovered with a fractional root-mean-square error (RMSE) of 0.16 relative to their catalog flux densities \citep{perley:butler:2017}. Since a real burst will not lie exactly at the declination of a holography source, we take a conservative fractional error of 0.2 for pulses within 20$^\circ$ in hour angle. Beyond 20$^\circ$ the beam model cannot be tested, as steady sources are too faint to be recovered from a single baseband snapshot, so we conservatively double the uncertainty to 0.4. Two further terms enter the intensity flux density calibration. The FFT-formed beam model $M_{formed}$ contributes a fractional RMSE of 0.17, measured by calibrating nine far-sidelobe Crab pulsar pulses with both intensity and baseband data, and the calibration constants in Equation \ref{eq:intensity_fluxcal} contribute a fraction RSME of 0.07, measured from 22 steady sources transiting the CHIME meridian. Adding these in quadrature with the holography term, the total fractional uncertainty on the intensity flux densities in this work is 0.27 within 20$^\circ$ in hour angle and 0.44 beyond it.

\section{Formed beam response of the brightest pulse}
\label{sec:beam_positions}
\begin{figure}
  \centering
  \includegraphics[width=0.6\textwidth]{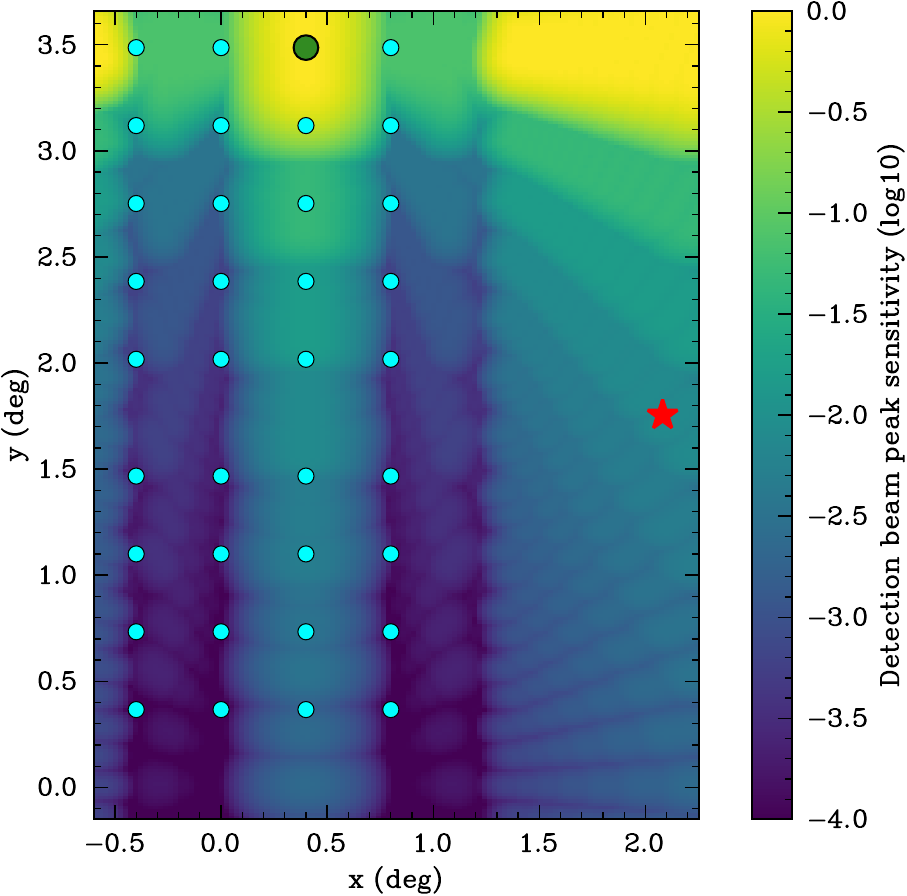}
  \caption{The formed beam response for the brightest detected pulse from \pulsar{}. The cyan circles are the CHIME/FRB formed beams, and the star marks the position of \pulsar{}. Unfortunately, for this pulse, data were saved only for a suboptimal beam, resulting in a noisier calibration, as is apparent in the left panel of Figure \ref{fig:brightest_burst}. The axis labels denote degrees from zenith in the east-west and north-south directions.}
  \label{fig:beam_positions}
\end{figure}
The formed beams in which the brightest pulse was recorded are shown in Figure \ref{fig:beam_positions}. We know that this pulse must originate from \pulsar{} because it follows other pulses with integer period separations.

\bibliography{main}{}
\bibliographystyle{aasjournalv7}
\end{document}